\documentclass[aps,prd,preprint,showpacs,superscriptaddress]{revtex4}
\usepackage{graphicx}
\usepackage{hyperref} 
\usepackage[usenames]{xcolor}

\usepackage[utf8x]{inputenc}
\usepackage{fancyhdr}
\usepackage{graphics}
\usepackage{epsf}
\usepackage{enumitem}
\usepackage{amsmath, amsthm, amssymb}
\usepackage{slashed}
\usepackage{subfig}
\usepackage{url}
\usepackage{ulem}
\usepackage{setspace}

\begin{document}

\title{Dark Matter Thermalization in Neutron Stars}
\author{Bridget Bertoni}
\email{bbertoni@uw.edu}
\affiliation{Department of Physics, University of Washington, Seattle, WA 98195}
\affiliation{Institute for Nuclear Theory, University of Washington, Seattle, WA 98195}
\author{Ann E. Nelson}
\email{anelson@phys.washington.edu}
\affiliation{Department of Physics, University of Washington, Seattle, WA 98195}
\author{Sanjay Reddy}
\email{sareddy@uw.edu}
\affiliation{Institute for Nuclear Theory, University of Washington, Seattle, WA 98195}


\begin{abstract}
We study how many-body effects alter the dark matter (DM) thermalization time inside neutron stars.  We find that Pauli blocking, kinematic constraints, and superfluidity and superconductivity in the neutron star significantly affect the DM thermalization time, in general lengthening it.  This could change the final DM mass and DM-nucleon cross section constraints by considering black hole formation in neutron stars due to DM accretion.  We consider the class of models in which DM is an asymmetric, complex scalar particle with a mass between 1 keV and 5 GeV which couples to regular matter via some heavy vector boson.  Interestingly, we find that the discovery of asymmetric, bosonic DM could motivate the existence of exotic neutron star cores.  We apply our results to the case of mixed sneutrino DM.
\end{abstract}

\maketitle


\section{Introduction}

Over the past 80 years there has been increasing evidence--from galactic rotation curves, the Bullet cluster, the anisotropies in the cosmic microwave background, the distribution of galaxies, etc.--for the existence of cold, non-baryonic dark matter (DM).  While many constraints have been placed on DM, its precise nature remains unknown (see \cite{bhs,dmreview,dmreviewfeng} for a review).  In particular, the mass of the DM particle $(m_{\chi})$ is highly unconstrained.  For elementary particle DM, we have that
\begin{equation}
10^{-22} \text{ eV} < m_{\chi} < 10^{19} \text{ GeV}~,
\end{equation}
where the lower bound comes from bosonic DM being confined on galaxy scales ($\lambda_{deBroglie} \sim$ kpc) and the upper bound is the Planck scale.  Phase space density bounds fermionic DM to be heavier than several keV \cite{Tremaine}.  Other constraints on DM apply, even for bosons, if the galaxy core/cusp problem is explained by warm DM \cite{warmDM}.  DM candidates with $m_{\chi} > 10^{19}$ GeV must be either black holes, e.g. primordial black holes \cite{PBHasDM,PBHconstraints,PBDMconstraints1,PBDMconstraints2}, or extended objects like composite DM or solitons, e.g. Q-balls \cite{QballDM,QballDMbounds}.  

Currently, direct detection experiments most strongly constrain the DM-nucleon cross section for DM with a mass $\sim$ 50 GeV.  The best constraint comes from XENON100 which requires the DM-nucleon cross section to be $\lesssim 10^{-45} \text{ cm}^2$ for $m_{\chi} \sim$ 50 GeV \cite{Xenon100}.  Lighter DM candidates with $m_{\chi} \lesssim 1 \text{ GeV}$ are less constrained by such experiments due to large backgrounds.  For this paper we will focus on the less-constrained parameter space of DM particles with mass between 1 keV and 5 GeV.  In this region, astrophysical observations can be used to constrain DM mass and cross section parameter space which is difficult for direct detection experiments to probe.

Such astrophysical observations were initially used to constrain hypothetical weakly interacting particles by studying their interactions inside of planets and stars \cite{ps1,ps2,gn}, and in the past few years there has been a renewed interest in using these methods to place constraints on DM by considering neutron star accretion of DM particles \cite{chargedDMNS,CompactStarsDM,WIMPanncool,NeutronStarsDM,progenitor,kt1,kt3,kt2,zurek,SelfInteracting,strongselfint,DMself1,DMself2}. Under certain conditions this accretion of DM results in a black hole that then destroys the neutron star.  Observationally we know that the oldest neutrons stars are $\sim$ 10 billion years old \cite{pulsarcatalog}, hence the DM parameter space which allows for neutron star destruction in less than 10 billion years is ruled out.

In what follows, we only consider bosonic DM since, due to the absence of Fermi pressure, it becomes gravitationally unstable for fewer accumulated particles than fermionic DM.  We also only consider asymmetric DM (for recent reviews see \cite{asymdmzurek,asymdm}), in which there is an initial asymmetry between particles and antiparticles so that today only particles remain and DM annihilation in the neutron star can be ignored.  Co-annihilation of DM with other particles in the neutron star was considered in \cite{DMself2} and will be neglected in our analysis.  Using black hole formation in neutron stars due to the accretion of asymmetric, bosonic DM, references \cite{kt1,kt3,kt2,zurek,NeutronStarsDM,strongselfint,DMself1,DMself2} have put otherwise model-independent constraints on DM.  Here we show that an improved treatment of particle kinematics and many-body effects including Pauli-blocking, superfluidity, and superconductivity can have a significant effect on the DM thermalization time, changing the final DM constraint.  We also show that in some phases of high density matter, such as color superconducting quark matter, thermalization times can be surprisingly large.  

This paper is organized as follows.  In section \ref{sec2}, we briefly review the nature of the DM constraint--the process of DM capture by a neutron star, black hole formation, and destruction of the neutron star.  In section \ref{sec3}, we discuss the effective theory for our generic DM model in light of this scenario.  In section \ref{sec4}, we define the thermalization time and compare our results to those previously obtained.  We conclude in section \ref{sec5} and discuss mixed sneutrino DM in appendix \ref{A} and \ref{B} as an example of a DM model which can be constrained in this way.


\section{Dark Matter Capture and Black Hole Formation}\label{sec2}

Here we review this process as previously discussed by many others (e.g. \cite{kt1,zurek}).  There are number of steps involved.  First, because of its gravitational interactions, DM is accreted by the neutron star.  We can estimate the velocity of the incident DM particle at the surface of the neutron star by using classical energy conservation:
\begin{equation}
\gamma(v) = \gamma(v_{\infty})+\frac{2GM}{R}~,
\end{equation}
where $\gamma(v) = (1-v^2)^{-1}$ and $v_{\infty}$ is the particle's velocity infinitely far from the neutron star.  We will take $v_{\infty} = 10^{-3}$.  For a typical neutron star, $M = 1.4M_{\odot} = 2.8 \times 10^{30}$ kg and $R = 10$ km.  Using these standard values, we find
\begin{equation}
v \simeq 0.7 ~.
\end{equation}
This implies that the energy that a typical DM particle has at the surface of a neutron star is
\begin{equation}
E = \sqrt{k^2+m_{\chi}^2} \simeq 1.4m_{\chi}~,
\end{equation}
so we see that the incident  DM energy is set by its mass and that typical DM particles are semi-relativistic.  These incident DM particles will scatter with quasi-particles inside the neutron star, lose energy, and become bound to the star.
  
Next DM thermalizes inside the neutron star.  Since the incident DM particle is at most semi-relativistic, and it must lose energy in order to be captured by the neutron star, it is safe to assume that the typical DM particle is non-relativistic during the latter collisions that determine its thermalization time.  As the DM thermalizes, it collects within a sphere of radius $r_{th}$ which satisfies
\begin{equation}\label{rth}
\frac{GM(r_{th})m_{\chi}}{r_{th}} \approx \frac{3}{2}T ~,
\end{equation}
where $M(r_{th})$ is the mass of the neutron star enclosed within  a radius $r_{th}$ and $T$ is the temperature of the neutron star.  We can estimate this by considering a neutron star with a constant core density $\rho_c = 0.5$ GeV/fm${}^3$ and we find \cite{progenitor}
\begin{equation}
r_{th} \approx 2.2 \text{ m } \left(\frac{T}{10^5 \text{ K}}\right)^{1/2}\left(\frac{\text{GeV}}{m_{\chi}}\right)^{1/2} ~.
\end{equation}
This tiny sphere of DM at the center of the neutron star can begin to self-gravitate and collapse into a black hole.  Gravitational collapse is accelerated if the captured DM forms a Bose-Einstein condensate inside the star \cite{kt3,alanBEC}.  Once the black hole is formed, it must be massive enough to avoid evaporation due to Hawking radiation and then it may consume the neutron star.  The observational signatures of a neutron star collapsing into a black hole is still an interesting, open question. 

In previous works, \cite{kt1,zurek}, two calculations to constrain the DM-neutron cross section as a function of DM mass are done: 1) the thermalization time calculation: $\tau_{therm} = 10^{10}$ years, in which $\tau_{therm}$ is the time necessary for DM thermalization with the neutron star and 2) an accretion time calculation: $\tau_{acc} = 10^{10}$ years, in which $\tau_{acc}$ is the time needed for the neutron star to accrete enough DM to form a black hole which will destroy the star, assuming thermalization occurs in a negligible amount of time.  The second calculation sets the final constraint, and the first is used to find regions where the second constraint is not valid.  In this paper we will consider only the first calculation and its application to the particular class of DM models to be discussed next.


\section{Dark Matter Model}\label{sec3}

We consider a model in which DM is a complex scalar particle which couples to regular matter by exchanging some heavy spin one boson.  The effective Lagrangian for the interaction between DM and the fermions (nucleons, electrons, etc.) that are found in neutron stars is then given by
\begin{equation}\label{Lagrangian}
\mathcal{L}_{int} = \tilde{G} \ell_{\mu} \left(j_V^{\mu} + \alpha j_A^{\mu}\right)~,
\end{equation}
where $\ell_{\mu} = \partial_{\mu}\chi^{\dagger}\chi - \chi^{\dagger}\partial_{\mu}\chi$ is the DM current, $j_V^{\mu} = \bar{\psi}\gamma^{\mu}\psi$ and $j_A^{\mu} = \bar{\psi}\gamma^{\mu}\gamma_5\psi$ are the vector and axial-vector currents for the fermions, and $\alpha$ is the coupling of $j_A^{\mu}$ to the mediator divided by the coupling of $j_V^{\mu}$ to the mediator.  For simplicity we take $\alpha$ to be the standard model value for fermions coupling to the Z boson.  $\tilde{G}$ is the coupling constant after the heavy mediator has been integrated out.  In general,
\begin{equation}\label{GF}
\tilde{G} = \frac{g_{\chi}g^V_{\psi}}{M_H^2}~,
\end{equation}
where $M_H$ is the mass of the heavy mediator particle, $g_{\chi}$ is the coupling of the mediator to $\ell^{\mu}$, and $g^V_{\psi}$ is the coupling of the mediator to $j_V^{\mu}$.

Use of this effective theory is well-justified.  In order for the effective theory to capture the relevant physics, one needs that the magnitude of the four-momentum transfer squared, $q^2$, is much less than $M_H^2$ in the DM-fermion scattering processes.  Based on the arguments in the previous section we know that the initial DM energy is at most $1.4m_{\chi}$ and hence the maximum $q^2$ that the DM can give up is $|q^2_{max,DM}| \approx 4m_{\chi}^2$.  Since the fermions inside the neutron star are highly degenerate, scattering events in which the DM gains energy and momentum from them are rare and have typical $\sqrt{q^2} \sim T \sim 9 \text{ eV} << m_{\chi}$ for the DM masses we are considering.  Note that we will always take $m_{\chi} \gtrsim 1$ keV as it was shown in \cite{kt3} that for $m_{\chi} \lesssim 1$ keV one must worry about captured DM escaping the star.  Hence as long as $m_{\chi} << M_{H}$, then $q^2 << M_{H}^2$ and our effective theory is valid.  For this reason we will take $m_{\chi} < 5$ GeV, consistent with the assumption that the heavy vector mediator is either a standard model $Z$ or $W^{\pm}$ pair, or a heavier, undiscovered particle.

Unless forbidden by some symmetry, this effective theory also includes DM self-interactions.  These have been shown to affect the critical number of DM particles needed for black hole formation \cite{DMself1,DMself2}, but are not relevant for thermalization time calculations and hence will not be discussed further here.   


\section{Thermalization Time Calculation and Results}\label{sec4}


\subsection{The Thermalization Time}

For scattering between DM and fermions, let $k^{\mu} = (E^{\chi}_k,\vec{k})$ be the initial DM four-momentum, $k'^{\mu} = (E^{\chi}_{k'},\vec{k'})$ be the final DM four-momentum, $p^{\mu} = (E^f_p,\vec{p})$ be the initial fermion four-momentum, and $p'^{\mu} = (E^f_{p'},\vec{p'})$ be the final fermion four-momentum.  We define the thermalization time as the average time it takes for an incident DM particle to start having collisions in which the average energy transfer is less than the temperature of the neutron star, i.e. $\langle q_0 \rangle \lesssim T$ where $q_0$ is the zeroth component of the four-momentum transfer $q_{\mu} = k_{\mu} - k'_{\mu}$.  Note that we will assume that the DM particles are confined to the neutron star interior.  While initially DM particles which have become bound to the neutron star may have an orbit that goes outside the star \cite{kt1}, unless DM is extremely light ($m_{\chi} \lesssim 1$ keV), the DM particles are confined to the neutron star interior for later stages of cooling.  Since we are considering the oldest, coldest neutron stars, we take $T = 10^5 \text{ K } \approx 9$ eV. 

To derive a formula for the thermalization time, we will make use of the DM scattering rate, $\Gamma$.  Using Fermi's Golden Rule, the scattering rate for DM scattering with a medium of spin $1/2$ fermions is given by 
\begin{align}\label{scattrate}
\Gamma = 2\int\frac{d^3p}{(2\pi)^3}\int\frac{d^3k'}{(2\pi)^32E^{\chi}_{k'}}\int\frac{d^3p'}{(2\pi)^32E^f_{p'}}(2\pi)^4\delta^4(p^{\mu}+k^{\mu}-p'^{\mu}-k'^{\mu}) \\
\times \frac{\langle|\mathcal{M}|^2\rangle}{2E^f_p2E^{\chi}_k}
n_F(E^f_p)\left(1-n_F(E^f_{p'})\right)\Big(1+n_B(E^{\chi}_{k'})\Big)~, \nonumber
\end{align}
where $\mathcal{M}$ is the amplitude for the process, $n_F$ is the fermion distibution function (Fermi-Dirac for non-interacting fermions), and $n_B$ is the DM distribution function (Bose-Einstein for non-interacting bosonic DM).  We will neglect the Bose enhancement factor for the final DM state for simplicity since the distribution function for the DM particles is a complicated function of time due to the accumulation of DM--note that this means that (\ref{scattrate}) as is, is actually a lower bound on the scattering rate.

The tree level squared amplitude, $\langle |\mathcal{M}|^2 \rangle$, is averaged over initial and summed over final fermion spins and is given by
\begin{align}\label{amplitude}
\langle|\mathcal{M}|^2\rangle=2\tilde{G}^2 & \left\{ (1+\alpha^2)  \left[2\left(p'\cdot(k+k')\right)\left(p\cdot(k+k')\right)-(p'\cdot p)(k+k')\cdot(k+k')\right] \right.  \\
 & ~ \left. + ~ (1-\alpha^2)\left[m_f^2(k+k')\cdot(k+k')\right]\right\}~,
\end{align}
where we have used the notation $a \cdot b \equiv a^{\mu}b_{\mu}$ with the mostly minus metric and $m_f$ is the fermion mass which could be the neutron or the electron mass, $m_n$ or $m_e$ respectively.  Using finite temperature formalism, the scattering rate can also be expressed as \cite{kapusta,greenfunc,NSneutrino}
\begin{equation}\label{scattrate2}
\Gamma = -2\tilde{G}^2 \frac{1}{1-e^{-q_0/T}}\int\frac{d^3k'}{(2\pi)^3} \frac{\text{Im}[\mathcal{L}^{\mu\nu}\Pi_{\mu\nu}^R]}{2E_k^{\chi}2E_{k'}^{\chi}}~,
\end{equation}
where $\mathcal{L}_{\mu\nu}$ contains the DM currents and $\Pi_{\mu\nu}^R$ is the fermion retarded polarization tensor.  For non-interacting fermions, these are given by
\begin{subequations}
\begin{align}
&\mathcal{L}_{\mu\nu} = (k+k')_{\mu}(k+k')_{\nu} ~\text{  and} \label{DMtensor} \\
\text{Im}\left[\Pi_{\mu\nu}^R\right] = \text{Im}\bigg[-i\tanh&\left(\frac{q_0}{2T}\right)& \nonumber \\ \times \int&\frac{d^4p}{(2\pi)^4}Tr[G(p)(\gamma_{\mu}+\alpha\gamma_{\mu}\gamma_5)G(p+q)(\gamma_{\nu}+\alpha\gamma_{\nu}\gamma_5)]\bigg] ~, \label{Ntensor}
\end{align}
\end{subequations}
where $G(p)$ is the free fermion propagator at finite temperature and density.  The form for this polarization tensor has been worked out in detail in \cite{NSneutrino} and \cite{japan} and we use their results in our calculations.  (If using the derivation in \cite{NSneutrino} note \cite{privcomm}.) 

The polarization tensor, $\Pi_{\mu \nu}^R$, characterizes the medium response to the DM probe.  The fermion propagators contain the Pauli blocking factors (c.f. the factor of $n_F(E_p^f)(1-n_F(E_{p'}^f))$ in (\ref{scattrate})) which restrict the fermion phase space due to the Pauli exclusion principle, i.e. the incident fermion that interacts with the DM particle must come from the initial fermion distribution and the scattered fermion must occupy phase space that is not already filled by the initial fermion distribution.  The polarization tensor also contains information about the in-medium fermion-fermion interactions since $\Pi_{\mu \nu}^R$ is a fermion current-current correlation function which includes a sum over all possible intermediate states.

Given an expression for the scattering rate ((\ref{scattrate}) or (\ref{scattrate2})), we can now define a discretized version of the thermalization time, $\tau$, based on the physical reasoning that the average thermalization time is simply the sum of the average times for subsequent DM collisions until the average energy transfer per collision is less than the temperature of the neutron star.  Thus we may write
\begin{equation}\label{numtau}
\tau = \frac{1}{\Gamma(E_0)} + \frac{1}{\Gamma(E_1)} + \frac{1}{\Gamma(E_2)} + \ldots + \frac{1}{\Gamma(E_n)} ~,
\end{equation}
where $E_0$ is the initial DM energy, which we will always estimate to be $1.05m_{\chi}$ (note that this assumes a $\sim 40 \%$ decrease in initial DM velocity due to prior collisions necessary for DM capture) and $E_i$ for $i > 0$ is the average final energy of a DM particle which had initial energy $E_{i-1}$.  The final energy $E_i$ is determined by calculating the scattering rate for a DM particle with initial energy $E_{i-1}$ weighted by the final DM energy, and dividing by the unweighted scattering rate for a DM particle with initial energy $E_{i-1}$, i.e.
\begin{equation}
\langle E_i(E_{i-1})\rangle = \frac{\int d\Gamma(E_{i-1})E^{\chi}_{k'}}{\int d\Gamma (E_{i-1})} ~.
\end{equation}
The summation in (\ref{numtau}) ends once $ \langle E_n - E_{n+1} \rangle < T$.  We expect that this generally results in $E_n \approx T$.  Expression (\ref{numtau}) is used for all of our numerical work.

We also define an approximate, continuous version of the thermalization time as
\begin{equation}\label{anatau}
\tau = -\int_{E_0}^{E_n}\frac{dE_i}{\int d\Gamma(E_i)(E_i-E_f)}~.
\end{equation}
We use (\ref{anatau}) for our analytic results, where instead of finding $E_n$ as described above, $E_n$ is fit to a value that approximates the numerical result well.

Now that the thermalization time is well defined, we can calculate the DM thermalization time inside a neutron star.  This includes DM scattering with a liquid of neutrons, protons, and electrons, a neutron superfluid, and a proton superconductor.  It also includes DM scattering with the matter in the core of the neutron star--possibly hyperons, pion or kaon condensates, quark gluon plasma, etc.  For a review of the constituents of a neutron star see \cite{glendenning,lattimerreview,reddyreview,neutronstar}.  The majority of the neutron star (roughly 85\%) is made up of neutrons, so we will first consider DM thermalization by scattering with neutrons, in both the normal (Fermi gas phase) and the superfluid phase.


\subsection{Scattering with a Fermi Gas of Neutrons}
\label{subsec:neutron}
From nucleon-nucleon scattering data we know that the neutron-neutron interaction can be either attractive or repulsive depending on the spin and spatial angular momentum of the neutrons and on the neutron density \cite{nnphaseshift}. At sufficiently low temperature, attractive interactions can lead to superfluidity and dramatically alter the low-lying excitation spectrum, and hence the DM scattering mechanism, and we discuss this in detail in the next section. Here, to calculate DM-neutron scattering, we will ignore nuclear interactions and approximate the neutrons as a dense, non-interacting Fermi gas. This will provide a baseline result since we know from Fermi-liquid theory that corrections due to strong  interactions in the normal phase do not qualitatively change the nature of scattering or the kinematics \cite{BCS}. From earlier work, relating to neutrino scattering in dense, normal neutron star matter \cite{BurrowsReddy}, we expect that the DM scattering rates in the Fermi gas approximation are sufficient to provide a reliable order of magnitude estimate. 

The fiducial calculation is done for neutrons at saturation density ($n_0\approx 0.16 \text{ fm}^{-3}$) which corresponds to a non-relativistic neutron chemical potential of $\mu_n \approx 0.056$ GeV.  This implies that neutrons at saturation density are to a good approximation, non-relativistic.  Deep in the core, neutrons become mildly relativistic and but these relativistic corrections are modest. We calculate the thermalization time and then enforce $\tau = 10^{10}$ years, which gives a constraint of the form $\tilde{G}$ as a function of $m_{\chi}$.  We then use this constrained coupling constant in the formula for the DM-fermion cross section in the limit in which both the DM and fermion momenta tend to zero (a good approximation of what takes place in direct detection experiments):
\begin{equation}\label{NRcrosssection}
\sigma_{DM-f} = \frac{\tilde{G}^2}{\pi}\frac{m_f^2m_{\chi}^2}{(m_f+m_{\chi})^2}~.
\end{equation}
This gives the DM-fermion cross section as a function of DM mass alone, with the constraint that DM thermalization takes longer than $10^{10}$ years.

For non-interacting neutrons it is simplest to use expression (\ref{scattrate}) for the scattering rate in the calculation of the thermalization time.  Eqn. (\ref{scattrate}) was used for numerical calculations and an approximate analytic result was obtained as follows.  For thermalization time scatterings it is a good approximation that both the neutrons and DM are non-relativistic, so neglecting all momentum dependence in the amplitude in (\ref{amplitude}) and rewriting the scattering rate we find
\begin{equation}\label{scattratecalc}
\Gamma \approx \tilde{G}^2\int\frac{d^3k'}{(2\pi)^3}S(q_0,q)~,
\end{equation} 
where $q_{\mu} = (q_0,\vec{q}) = k_{\mu}-k'_{\mu}$ is the four-momentum transfer and $q = |\vec{q}|$.  $S(q_0,q)$ is the neutron response function, here given by
\begin{equation}
S(q_0,q) = 2\int\frac{d^3p}{(2\pi)^3}\int\frac{d^3p'}{(2\pi)^3}(2\pi)^4\delta^4(p^{\mu}+k^{\mu}-p'^{\mu}-k'^{\mu})n_F(E^n_p)\left(1-n_F(E^n_{p'})\right)~,
\end{equation}
where $n_F(E) = [1+e^{(E-\mu)/T}]^{-1}$ is the Fermi-Dirac distribution function.  Additionally in the limit of completely degenerate neutron matter (in reality $\mu_n/T \sim 6.5 \times 10^6$ so the neutrons really are quite degenerate) and for $q << m_n$, we have \cite{NSneutrino}:
\begin{equation}\label{responsefunc}
S(q_0,q) \approx \frac{m_n^2T}{\pi q}\left(\frac{z}{1-e^{-z}}\right)\Theta(qv_F-|q_0|)~,
\end{equation}
where $z = q_0/T$, $\Theta$ is the Heaviside step function, and $v_F = p_F/m_n \approx 0.35$ is the neutron Fermi velocity.  

\begin{figure}[h!]
\includegraphics[scale=0.52]{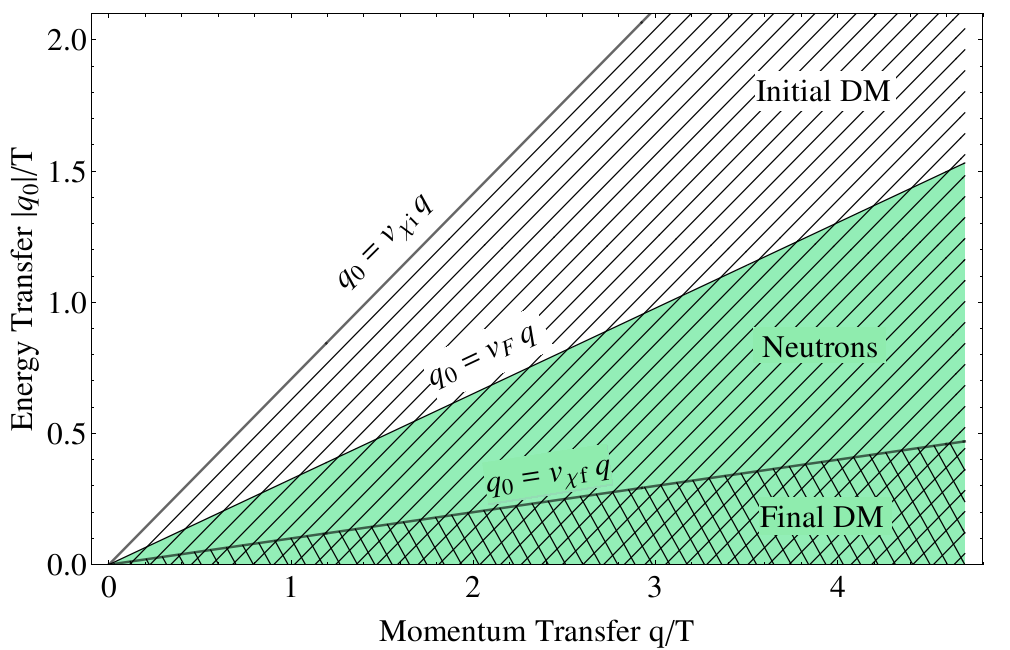}
\caption{Plot of the magnitude of the momentum transfer as a function of energy transfer, both in units of T, for momentum transfers much less than the mass and momentum of the particle involved.  The shaded areas show kinematically allowed regions.  The positively sloped lined region is for initial DM (with $v_{\chi i} = 0.7$), the green, shaded region is for neutrons, and the negatively sloped lined region is for final DM (with $v_{\chi f} << v_{\chi i}$ after the DM has lost energy to the neutrons).  DM-neutron scattering can take place in the kinematic regions where the DM and neutron regions overlap.}
\label{kinematics}
\end{figure}

Note that the step function is just enforcing non-relativistic, low momentum transfer neutron kinematics, i.e. that $|q_0| < v_Fq$.  That this inequality holds can be seen simply from
\begin{equation}
q_0 = E^n_{p'}-E^n_p = \sqrt{m_n^2-(\vec{p}+\vec{q})^2} - \sqrt{m_n^2+p^2} = \frac{pq\cos\theta}{E^n_p} + \mathcal{O}\left(\frac{q^2}{E_n}\right)~, 
\end{equation}
where $\theta$ is the angle between $\vec{p}$ and $\vec{q}$.  These neutron kinematics must be consistent with the same non-relativistic, low momentum transfer DM kinematics ($|q_0| < v_{\chi}q$) and since $v_{\chi} \leq 1/3$ always by construction, the DM kinematics constrain the phase space more and the neutron step function in (\ref{responsefunc}) can simply be set to 1.  These kinematics are shown in Fig. \ref{kinematics}.

\begin{figure}[h]
\includegraphics[scale=0.52]{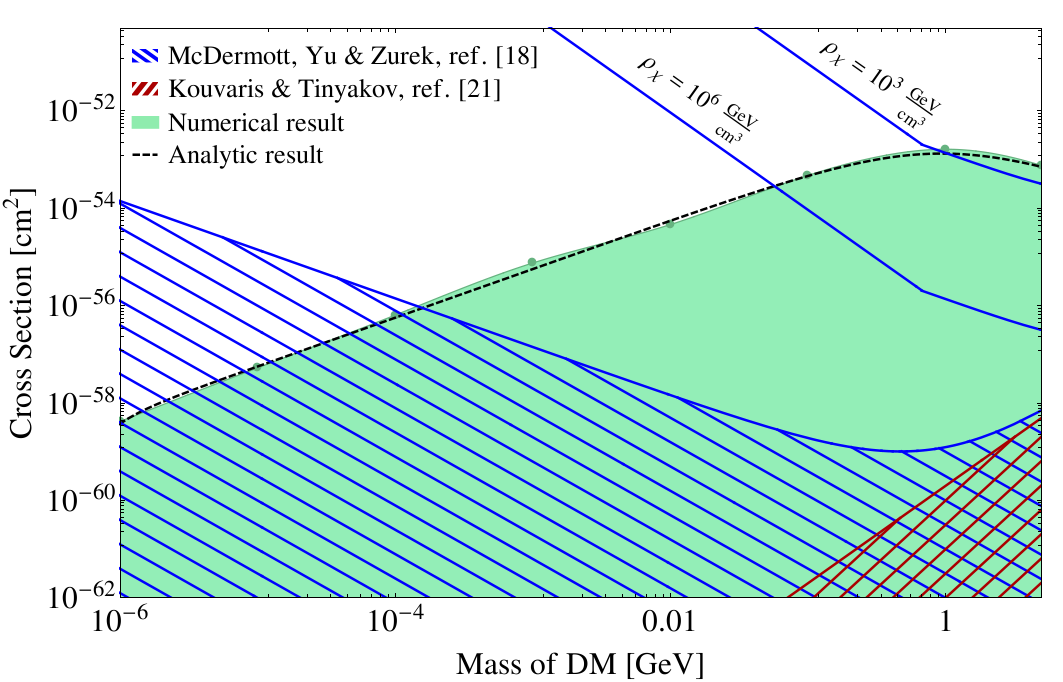}
\caption{
Plot of the DM-neutron cross section for DM interacting with a Fermi gas of neutrons.  Shaded regions are where DM takes longer than $10^{10}$ years to thermalize.  Lines labeled with different values of $\rho_{\chi}$ (the DM density around a neutron star--note \cite{DMdensity}) indicate upper bounds on the allowed DM-neutron cross section due to neutron stars accreting enough DM to form a black hole as computed in \cite{zurek} in the absence of DM self-interactions.}
\label{neutron_constraints}
\end{figure}

Using (\ref{responsefunc}) in (\ref{scattratecalc}), setting $e^{-z}$ to zero as the thermalization time definition always has $q_0 > T$ and completing the angular integrals gives
\begin{equation}
d\Gamma \approx \frac{\tilde{G}^2m_n^2}{4\pi^3} k'^2 q_0\left(\frac{k+k'-|k'-k|}{kk'}\right) dk'~.
\end{equation}
Since the neutrons are approximated as completely degenerate, DM cannot lose energy to them, hence $k' \leq k$, and using $q_0 = k^2/(2m_{\chi}) - k'^2/(2m_{\chi})$, we find
\begin{equation}\label{dGamma}
d\Gamma \approx \frac{\tilde{G}^2m_n^2}{2\pi^3k}k'^2 \left(\frac{k^2}{2m_{\chi}}-\frac{k'^2}{2m_{\chi}}\right) dk'~.
\end{equation}
We can now use this to calculate the denominator in (\ref{anatau}):
\begin{equation}
\int d\Gamma(E_i)(E_i-E_f) = \int d\Gamma(E_k^{\chi})(E_k^{\chi}-E_{k'}^{\chi}) \approx \frac{\tilde{G}^2m_n^2}{2\pi^3k}\int_0^k k'^2 \left(\frac{k^2}{2m_{\chi}}-\frac{k'^2}{2m_{\chi}}\right)^2 dk'~.
\end{equation}
Integration gives
\begin{equation}
\int d\Gamma(E_i)(E_i-E_f) \approx \frac{\tilde{G}^2m_n^2}{105\pi^3m_{\chi}^2}k^6~.
\end{equation}
Using this in (\ref{anatau}) we find
\begin{equation}\label{tau}
\tau \approx \frac{105\pi^3m_{\chi}}{4\tilde{G}^2m_n^2}\left(\frac{1}{k_n^4}-\frac{1}{k_0^4}\right)~.
\end{equation}
Setting $k_0 = m_{\chi}/3$, using $k_n = \sqrt{4m_{\chi}T}$ to match to numerical calculations, and enforcing $\tau \geq 10^{10}$ years gives the final result for $\tilde{G}(m_{\chi})$.  Our numerical and analytic results are shown in Fig. \ref{neutron_constraints} along with previous results for comparison.  Note that the result shown from \cite{zurek} is their full analytic result and not the approximation that they plot in their figures.

In order to compare with analytic expressions from previous works \cite{zurek,progenitor}, we neglect $k_0^{-4}$ with respect to $k_n^{-4}$ in (\ref{tau}) and insert (\ref{NRcrosssection}) into the expression to obtain
\begin{equation}
\tau \approx \frac{105\pi^2}{16m_n\sigma T}\frac{\gamma}{(1+\gamma)^2}~,
\end{equation}
where $\gamma \equiv m_{\chi}/m_n$.  To get a feel for typical scales, this can be recast as
\begin{equation}
\tau \approx 3750 \text{ yrs }\frac{\gamma}{(1+\gamma)^2}\left(\frac{2 \times 10^{-45} \text{ cm}^2}{\sigma}\right)\left(\frac{10^5 \text{ K}}{T}\right)^2~,
\end{equation}
which is generically longer than previous calculations by several orders of magnitude.

\begin{figure}[h!]
\includegraphics[scale=0.46]{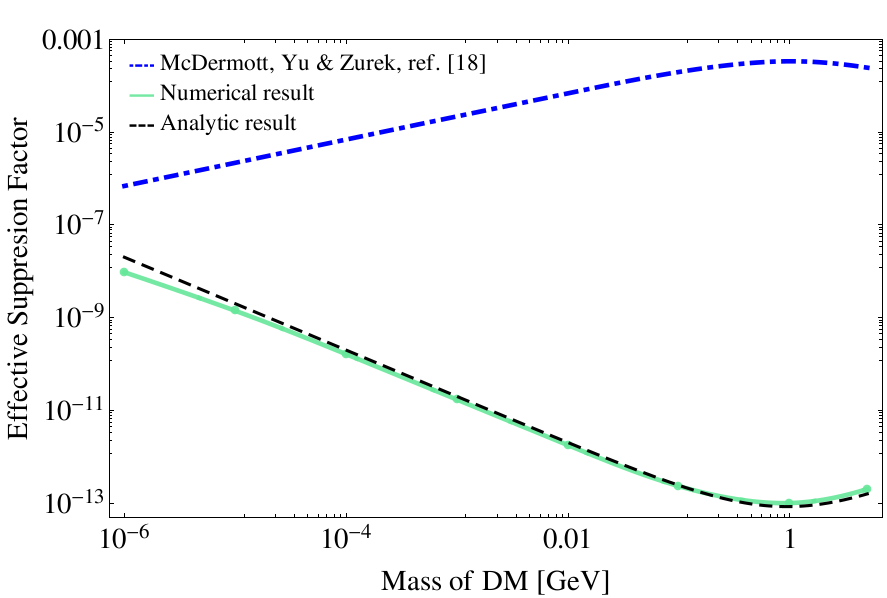}
\caption{Comparison of effective suppression factors in the DM-neutron scattering rate for roughly thermal DM ($E = 3T$) and a non-interacting Fermi gas of neutrons.}
\label{pauliblocking}
\end{figure}

From Fig. \ref{neutron_constraints} one can see that the results obtained here differ appreciably from those in previous works--in particular some regions of DM parameter space that were disallowed in \cite{zurek} are allowed from this calculation due to an increase in thermalization times.  This is because the proper inclusion of kinematics and Pauli blocking are essential for calculating the low energy and momentum transfer scattering processes that lead to thermalization.  In the past Pauli blocking has been included only roughly, and kinematic effects have been neglected.  
To estimate the contribution of Pauli blocking and kinematic constraints to our calculations we define an effective suppression factor $\xi_{eff}$, given by
\begin{equation}
\xi_{eff} = \frac{\Gamma}{n \sigma v}~,
\end{equation}
where $n \sigma v$ is the classical expression for the scattering rate, $n = p_F^3/(3\pi^2)$ is the number density of the neutrons, $\sigma$ is the cross section given in (\ref{NRcrosssection}), and $v$ is the magnitude of the relative velocity between the incident DM and incident neutron. $\Gamma$ (see (\ref{scattrate})) is the actual scattering rate which is an integrated version of $n \sigma v$.
Using (\ref{NRcrosssection}), (\ref{dGamma}) after integrating over $k'$, and using a thermal $k = \sqrt{6m_{\chi}T}$ for the incident DM momentum, we find
\begin{equation}
\xi_{eff} \approx \frac{18T^2(m_{\chi}+m_n)^2}{5k_F^3m_{\chi}\left\lvert\sqrt{\frac{6T}{m_{\chi}}}-\frac{k_F}{m_n}\right\lvert}~.
\end{equation}
In Fig. \ref{pauliblocking} we compare our suppression factor to the Pauli blocking suppression factor used in \cite{zurek}.
Fig. \ref{pauliblocking} shows that the inclusion of Pauli blocking and kinematic effects in the properly integrated scattering rate makes a difference.  These scattering kinematics and Pauli blocking are unique to DM scattering with a non-interacting, non-relativistic Fermi gas and will change once interactions between the fermions are included, especially in the case of attractive interactions which can give rise to superfluidity or superconductivity.  We discuss these effects next.


\subsection{Scattering with Superfluid Neutrons}

From BCS theory we know that attractive interactions at the Fermi surface leads to the formation of Cooper pairs at low temperature and results in a phase transition to either a superfluid or superconducting state \cite{BCS}. In this superfluid or superconducting state, there is a non-zero ground state expectation value (or condensate) of Cooper pairs which produces a gap in the fermion excitation spectrum and a Goldstone boson due to the spontaneous breaking of the $U(1)$ symmetry associated with fermion number \cite{BCS, Weinberg}.

For neutrons in the core of the neutron star the dominant attractive interaction is in the p-wave channel and is expected to lead to the formation of spin-triplet Cooper pairs \cite{neutronstar}. Model calculations predict that the energy gap, $\Delta_{\rm {}^3P_2}$, is roughly $0.01-0.1$ MeV, though this remains somewhat uncertain \cite{SchwenkBengt}. The condensate of these pairs is expected to be spatially anisotropic and Goldstone bosons associated with the breaking of rotational invariance arise in addition to the Goldstone boson from the spontaneous breaking of fermion number \cite{Savage}. Since in our model, DM couples only to the neutron density in the non-relativistic limit (in (\ref{DMtensor}), $\mathcal{L}_{\mu \nu} \rightarrow 4m_{\chi}^2\delta_{\mu 0}\delta_{\nu 0}$), the only relevant excitation at energies small compared to the gap is the Goldstone boson, or superfluid phonon, associated the breaking of the $U(1)$ fermion number symmetry. 


The superfluid phonon manifests as spikes in the density-density neutron response function ($\sim \text{Im}[\Pi_{00}^R]$, c.f. (\ref{Ntensor})) at $|q_0| = c_s q$, where $c_s$ is the speed of the superfluid phonon in the nuclear medium.  Based on this, we can make an ansatz for the neutron response function:
\begin{equation}\label{1phononS}
S(q_0,q) = A \delta(q_0-c_sq) + B\delta(q_0+c_sq)~,
\end{equation}
where $A$ and $B$ are normalization constants which can be fixed by enforcing the principle of detailed balance and the f-sum rule \cite{plischke}
\begin{equation}
S(q_0,q) = e^{q_0/T}S(-q_0,q)~~\text{  and  }~~\frac{1}{2\pi}\int_{-\infty}^{\infty} dq_0 q_0(1-e^{-q_0/T})S(q_0,q) = \frac{q^2}{m_n}n~ .
\end{equation}
This gives
\begin{equation}
S(q_0,q) = \frac{\pi n q}{m_nc_s}\left[\frac{\delta(q_0-c_sq)}{1-e^{-q_0/T}}+\frac{\delta(q_0+c_sq)}{e^{-q_0/T}-1}\right]~ .
\end{equation}

In our calculations we will only use the part of $S$ $\propto \delta(q_0-c_sq)$ which allows the DM to lose energy.  The $\delta(q_0+c_sq)$ part is also suppressed for thermalization scatterings with $|q_0| > T$.  We take
\begin{equation}
S(q_0,q) \approx \frac{\pi n q}{m_nc_s}\left[\frac{\delta(q_0-c_sq)}{1-e^{-q_0/T}}\right]~.
\end{equation}
Note that this response function only characterizes DM emission of a single phonon.  Multi-phonon processes are quite suppressed and will be discussed in section \ref{subsec:exotic}.  This response function can be used in place of the neutron part of the scattering rate in (\ref{scattrate}) and then the DM thermalization time in a neutron superfluid can be computed.  In doing so, we varied $c_s$ between $0.5v_F/\sqrt{3}$ and $3v_F/\sqrt{3}$.  The value $v_F/\sqrt{3}$ is the leading order speed of the superfluid phonon \cite{SFLeff}. 
In general $c_s = \sqrt{\partial P / \partial \rho}$, where $P$ is the pressure in the nuclear medium and $\rho$ is the energy density; $c_s$ is expected to vary inside the neutron star due to phonon interactions and as a function of density.

\begin{figure}
\centerline{\includegraphics[scale=0.405]{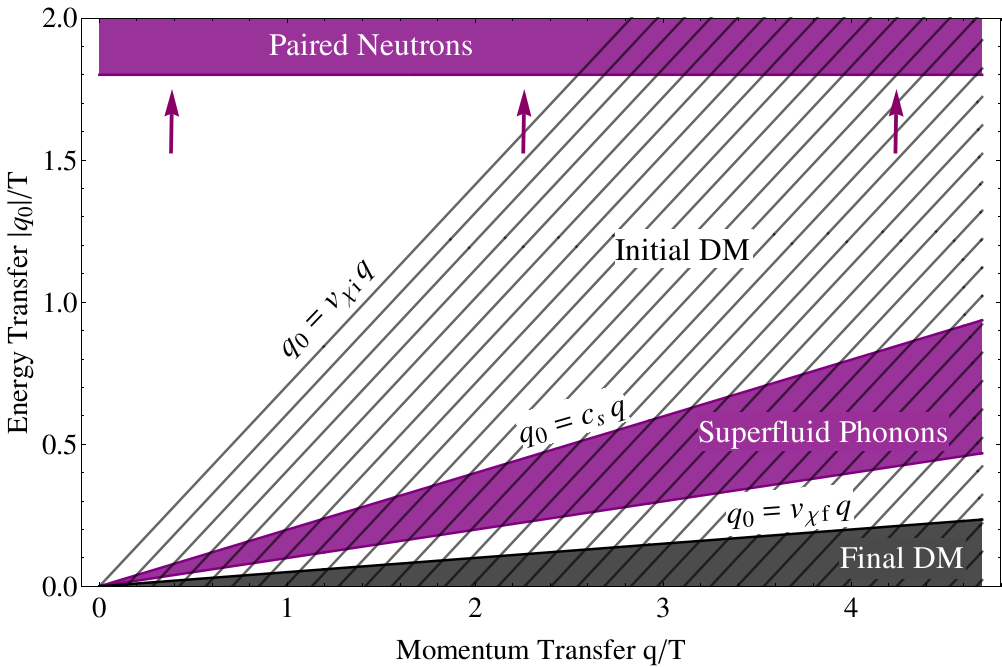}}
\caption{Plot of DM and neutron superfluid kinematically allowed regions.  The neutron superfluid region includes a kinematic region for the phonon mode as well as one for neutron-pair interactions which begins at $ q_0/T \sim 10^3$ on this scale.  If DM is travelling at a speed larger than the speed of the superfluid phonon, the DM and neutron superfluid kinematic regions overlap and scattering can occur.  However after DM scatters and loses energy, its speed decreases and the DM and neutron superfluid kinematic regions no longer overlap and no more scattering can occur.}
\label{SFkinematics}
\end{figure}

We find in general that DM particles will only scatter with the neutron superfluid once or twice, leaving the DM with too much energy to be considered thermal.  This result is due to the highly restricted kinematics of the neutron superfluid, see Fig. \ref{SFkinematics}.  Since the single phonon mode can only respond with $q_0 = c_s q$, once the DM particle loses enough energy such that $v_{\chi} < c_s$, neutron superfluid and DM kinematics are no longer compatible and no further scattering can occur.

Since DM cannot thermalize by single phonon emission in the neutron superfluid, the DM particle must scatter with something else inside the neutron star in order to thermalize.  In addition to suppressed multi-phonon scattering, the other standard options are scattering with protons or electrons.  The protons are likely to be in a superconducting state with Cooper-paired protons and a massless, coupled proton and electron mode \cite{protonGB}.  The paired protons have a gapped energy spectrum and hence do not contribute much to DM thermalization.  The massless proton-electron mode has kinematics similar to that of the neutron superfluid phonon mode, so it also does not allow DM to thermalize.  This only leaves the electrons to thermalize the DM.


\subsection{Scattering with a Fermi Gas of Electrons}
\label{subsec:electron}
Electrons in a neutron star have a vanishingly small critical temperature for pairing, so the low-energy spectrum of particle-hole excitations is un-gapped and well described by that of a non-interacting Fermi gas.  Thus electron-DM scattering can be treated in the same way as the neutron-DM scattering in section \ref{subsec:neutron}.  Roughly $7\%$ of a neutron star is made up of electrons and for neutrons at saturation density, electrons have a chemical potential of $\mu_e \approx 0.12$ GeV, indicating that the electrons are highly relativistic with the Fermi velocity $v_F \approx 1$.  Since DM is non-relativistic, its dominant coupling is to the electron density but the kinematics differs qualitatively from the neutron case because $v_F \approx 1$ and DM-electron scattering is always kinematically allowed inside the neutron star.  

\begin{figure}[h!]
\includegraphics[scale=0.42]{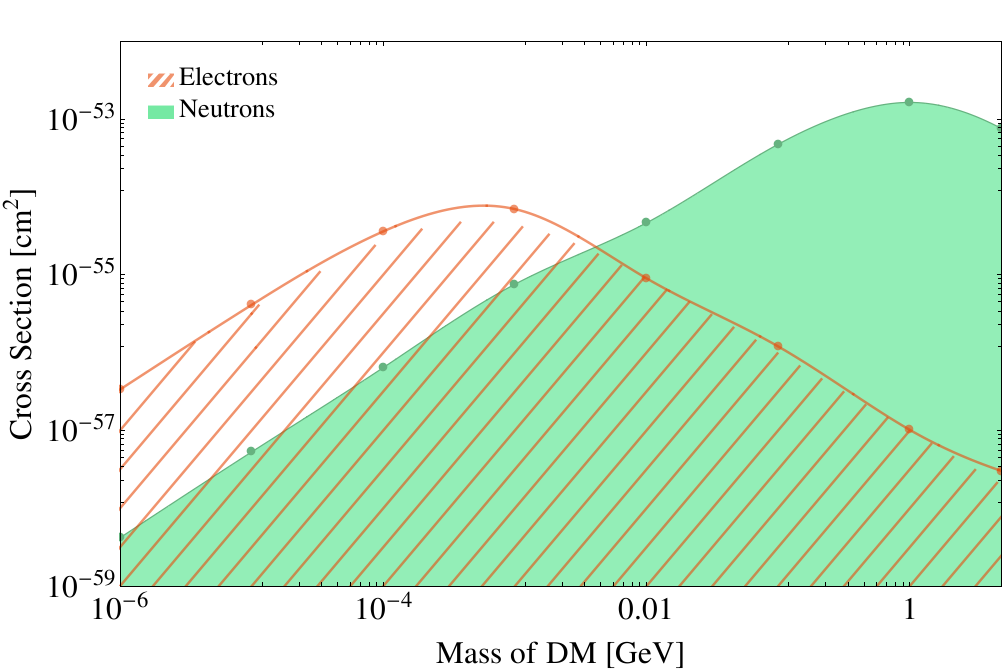}
\caption{Plot of the low energy DM-neutron and DM-electron cross sections as a function of DM mass.  Shaded areas are regions where DM thermalization takes longer than $10^{10}$ years.}
\label{enplot}
\end{figure}

The electron response function to leading order in the velocity of the DM particle is given by 
\begin{align}
S(q_0,q) = \int\frac{d^3p}{(2\pi)^3}\int\frac{d^3p'}{(2\pi)^3}\bigg[(2\pi)^4\delta^4(p^{\mu}+k^{\mu}-&p'^{\mu}-k'^{\mu})(1+\cos\theta) \\
\times &~n_F(E^n_p)\left(1-n_F(E^n_{p'})\right)\bigg]~,
\end{align}
where $n_F(E) = [1+e^{(E-\mu)/T}]^{-1}$ is the electron Fermi-Dirac distribution function and $\theta$ is the angle between $\vec{p}$ and $\vec{p}'$.  In Fig. \ref{enplot} we show the numerical results obtained from setting $\tau \geq 10^{10}$ years (using (\ref{scattrate2}),(\ref{numtau}), and (\ref{NRcrosssection})) for the low energy DM-electron cross section as a function of DM mass. The DM-neutron cross section results are plotted for comparison.
Interestingly, if DM couples with equal strength to neutrons and electrons (i.e. if $\tilde{G}$ is fixed), then we find that thermalization times for DM scattering with electrons are roughly 50\% of thermalization times for DM scattering with neutrons, so regardless of the presence of a superfluid, DM-electron scattering would be the most efficient process for DM thermalization.


\subsection{Scattering in Exotic Neutron Star Cores}
\label{subsec:exotic}

So far we have considered DM thermalization with electrons and also neutrons, both in the normal phase and in the superfluid phase.  However, the phase structure of matter in the neutron star core remains uncertain \cite{reddyreview}.  In this section we study two specific phases of high density matter in order to explore their influence on DM thermalization.  At asymptotically large densities where the strange quark mass can be considered small and perturbation theory is applicable, it is now well established on theoretical grounds that the ground state of quark matter is the color flavor locked (CFL) phase in which the $SU(3)_C \times SU(3)_L \times SU(3)_R \times U(1)_B$ approximate symmetry of QCD is spontaneously broken down to its vector subgroup $SU(3)_{C+L+R}$ due to the formation of a condensate of di-quark pairs \cite{colorSC1,colorSC2}. This is a color superconducting phase in which all nine (3 flavors $\times$ 3 colors) light quarks form Cooper pairs and there is a gap in the particle-hole excitation spectrum.  

At densities of relevance to neutron stars, the strange quark mass is dynamically important, perturbation theory fails, and whether the CFL phase is present at these densities remains an open question. If it were present, the CFL phase could additionally contain a condensate of K${}^0$ mesons \cite{BedaqueSchafer,KaplanReddy}. Both the CFL and the CFLK${}^0$ phases are characterized by similar low energy properties at temperatures of relevance to old neutron stars. They are both devoid of electrons and the only relevant low energy degrees of freedom are the massless Goldstone bosons associated with the breaking of global symmetries in the ground state \cite{colorSC1,colorSC2}. There is one massless phonon mode, sometimes called the $h$ boson, due to the breaking of the $U(1)_B$ symmetry in the CFL phase and two massless phonon modes in the CFLK${}^0$ phase, one due to the breaking of $U(1)_B$ (called $h$) and another due to the breaking of the hypercharge symmetry by the $K^0$ condensate (called $K_1$). The velocity of the $h$ mode in the relativistic limit is approximately given by $c_h \simeq 1/\sqrt{3}$ and the velocity of the $K_1$ is $c_{K_1} \simeq \sin{\theta} /\sqrt{3+9\cos^2{\theta}}$ where $ \sin{\theta}$ is proportional to the number density of the kaon condensate \cite{KaplanReddy}.

\begin{figure}[t]
\centerline{\includegraphics[scale=0.43]{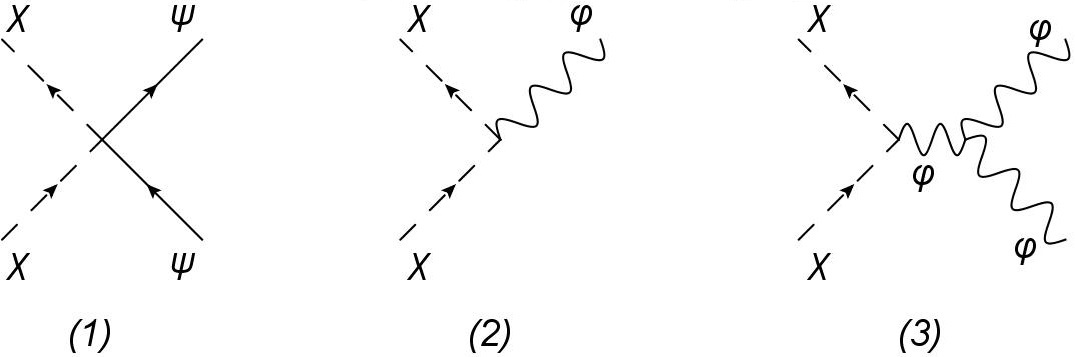}}
\caption{Scattering processes which contribute to DM thermalization inside a neutron star. $\chi$ denotes DM, $\psi$ is a neutron or electron, and $\phi$ is a superfluid phonon. (1) shows DM scattering with a non-interacting neutron or electron, (2) is DM scattering by emission of a single phonon, and (3) shows DM scattering a thermal phonon.}
\label{DMinteractions}
\end{figure}

Earlier we found that the gap in the nucleon spectrum due to pairing implied that DM thermalization would proceed via superfluid phonon emission processes as long as $v_{\chi} > c_s$, where $c_s$ was the speed of the Goldstone mode in the nuclear medium. For $v_{\chi} < c_s$, this process is kinematically forbidden and electron scattering dominates. In the CFL and CFLK${}^0$ phases electrons are absent and relevant DM thermalization processes can only involve the massless Goldstone bosons. As in the superfluid nuclear phase, thermalization in the CFL and CFLK${}^0$ phases proceeds by the phonon emission process shown by the second diagram in Fig. \ref{DMinteractions} (akin to the Cherenkov radiation of fast particles) as long as $v_{\chi} < c_h$ if DM couples to the baryon number current and $v_{\chi} < c_{K_1}$ if it also couples to the hypercharge current.  When $v_{\chi} < c_h$, DM thermalization cannot proceed by phonon emission and the dominant thermalization process is the two phonon process shown by the third diagram in Fig. \ref{DMinteractions}. Here, the initial state phonon is thermal with energy $p_0 = c_hp \sim T$, and the intermediate phonon is off-shell. 

For simplicity we will consider DM that couples only to the the baryon number. In this case, the low energy effective Lagrangian  has the form $ {\mathcal L}_{eff} = \tilde{G} f_h l^0(\partial_0 \phi - c_h \partial_i \phi)$ where $\phi $ is half of the overall phase of the condensate that breaks the $U(1)_B$ symmetry and $f_h \simeq \mu_q$ where $\mu_q \simeq 400$  MeV is the quark chemical potential in the neutron star core \cite{CFL_EFT}.  Then the amplitude for the DM + phonon $\rightarrow$ DM + phonon process can be approximated by
\begin{equation} 
{\mathcal M} \approx -4i\frac{\tilde{G} c_3 m_{\chi}^2}{f_h} ~\frac{q_0^2p_0p'_0}{c_h^2 q^2}~,
\label{eq:M}
\end{equation}  
where $c_3$ is the dimensionless constant that sets the strength of the leading order three phonon vertex $\simeq c_3 ~(\partial_0 \phi)^3/f_h^2$ in the low energy theory, $q_0$ is the energy of the intermediate phonon, $q$ is the magnitude of the momentum of the intermediate phonon, and $p_0$ and $p'_0$ are the initial and final phonon energies respectively.  We used an approximate form of the phonon propagator, $\sim (q_0^2-c_h^2q^2)^{-1} \approx -(c_h^2q^2)^{-1}$, since the phonon must be off-shell with $|q_0| < c_h q$. Using (\ref{eq:M}) the scattering rate can be calculated for arbitrary initial DM velocities and to first order in $v_\chi/c_h$. Taking typical values for our parameters: $p_0 \sim T$, $p'_0 \sim T$, $q_0 \sim \frac{v_{\chi}}{c_h}T$, and $q \sim T/c_h$,  we find that the dimensional estimate (ignoring factors of $2$ and $\pi$) for the DM-phonon scattering rate is given by        
\begin{equation}\label{eq:gammacfl} 
\Gamma \approx \left(\frac{\tilde{G}c_3}{f_h}\right)^2\frac{v_{\chi}^3T^7}{c_h^6}~.
\end{equation} 
Using $c_h \sim 1/\sqrt{3}$, $c_3 \sim 1$ and $v_{\chi} \sim 1/3$ in (\ref{eq:gammacfl}) for a single DM-phonon scattering process, and estimating $\tau \approx 1/\Gamma$ we find that the DM thermalization time is approximately 
\begin{equation}
\tau \approx 9 \times 10^{38} \text{ yrs } \left(\frac{\text{GeV}^2}{\tilde{G}}\right)^2\left(\frac{10^5 \text{ K}}{T}\right)^7~,
\end{equation} 
indicating that the DM scattering rate is too low to allow for thermalization even for the oldest neutron stars with ages $\sim 10^{10}$ years if the core contains either the CFL or CFLK$^{0}$ phase.


\section{Conclusions}\label{sec5}

We considered a relatively generic DM model in which the DM particle was a complex scalar that coupled to regular matter via some heavy vector boson, with the regular matter vector and axial-vector couplings to the heavy mediator taken to be those to the Z boson.  We then calculated DM thermalization times inside a neutron star for DM scattering with electrons, neutrons in the normal phase, neutrons in the superfluid phase, and color superconducting quarks.  

We found several important results:
\begin{itemize}
\item Including kinematics in the thermalization time calculation resulted in DM thermalization times that were qualitatively different from past results.
\item Previously neglected DM-electron scattering in ordinary neutron star cores is actually quite important.  It is a more efficent DM thermalization mechanism than DM-neutron scattering when the neutrons are in a non-superfluid state (for a fixed $\tilde{G}$) and it is the only relevant DM thermalization mechanism when the neutrons form a superfluid and the protons form a superconductor.
\item Exotic neutron star cores with color superconducting quark matter and no electrons give rise to very large thermalization times which protects neutron stars from their possible destruction as a result of DM accretion.  Hence the discovery of asymmetric, bosonic DM could motivate the existence of exotic neutron star cores.
\end{itemize}


\begin{acknowledgments}
This work was supported in part by the U.S. Department of Energy Grants DE-FG02-00ER41132 (SR and BB) and DE-FG02-96ER40956 (AEN and BB).  BB would also like to acknowledge partial support from an ARCS Foundation Fellowship. 
\end{acknowledgments}


\appendix 
\section{Non-thermal Dark Matter} \label{A}

Light DM candidates such as mixed sneutrinos can be produced non-thermally via the Affleck-Dine mechanism.  The Affleck-Dine mechanism was originally associated with baryogensis \cite{admech} but it can be applied to any scalar field that can have a large vev and whose interactions are negligible.  In a cosmological context, the end result of this mechanism is to non-thermally produce a large number of nearly zero-momentum particles--exactly what is needed for cold DM.  

To see this explicitly, consider the Lagrangian density for a complex scalar field $\chi$ with mass $m_{\chi}$ in curved space time with metric $g$,
\begin{equation}
\mathcal{L} = \sqrt{-\text{det }g}\left(|\partial_{\mu}\chi|^2 - m_{\chi}^2|\chi|^2 - V_{int}(\chi)\right)~.
\end{equation}
In the following we neglect the interaction terms, i.e. we take $V_{int} \rightarrow 0$, and we assume that inflation smooths out any spatial dependence in our field so that $\chi \approx \chi(t)$.  Working under the assumption that the universe is flat, isotropic, and homogeneous, we use the Friedmann-Robertson-Walker metric with scale factor $a(t)$.  Thus the classical equation of motion for our field is
\begin{equation}
\ddot{\chi} + 3H(t)\dot{\chi} + m_{\chi}^2\chi = 0~,
\end{equation}
where dots indicate time derivatives and $H(t) = \dot{a}/a$ is the Hubble parameter.

This is a damped harmonic oscillator equation, so that for $H(t) >> m_{\chi}$ (which occurs early in the universe) $\chi$ is overdamped and hence is frozen at its initial vev, $\chi(t) = \chi_0$.  For late times with $m_{\chi} >> H(t)$, $\chi$ oscillates (only in time) with its natural frequency $m_{\chi}$ and we find that the energy density of the field, $\rho$, is proportional to $\chi_0$ and scales like $a^{-3}$ just like the energy density for highly non-relativistic matter:
\begin{equation}
\rho \propto m_{\chi}^2\chi_0^2\left(\frac{1}{a}\right)^3~.
\end{equation}
For very non-relativistic matter, $\rho = mn$, where $n$ is the number density of particles, thus we also have that $n \propto m_{\chi}\chi_0^2/a^3$ so that for the proper choice of the initial vev $\chi_0$, we can match $\chi$'s number density today to what we observe for DM, i.e.
\begin{equation}
n(t_0) = \frac{1}{m_{\chi}}\left(\frac{3H_0^2}{8\pi G}\right)\Omega_{CDM}~,
\end{equation}
where $t_0$ is the time today, $G$ is the gravitational constant, $H_0$ is the Hubble parameter today, and $\Omega_{CDM} \approx 0.32$ \cite{planck}.  We also note that the DM masses considered in this paper are large enough so that to match the energy density today, the vev $\chi_0$ is small enough such that DM is present (beginning roughly at the time that satisfies $H(t)=m_{\chi}$) before matter domination, as required by observations of the CMB.

It is easy to explain why interaction terms for $\chi$ are negligible, as well as why a macroscopically large vev can form during the early universe if $\chi$ can be non-zero along a flat direction in the scalar potential.  Such flat directions are common in supersymmetric theories; for example, it is possible to find flat directions with combinations of squarks, sleptons, and Higgs fields in the Minimal Supersymmetric Standard Model \cite{flatdirections}. 

\section{Light Sneutrino Dark Matter}\label{B}
In this appendix we consider Affleck-Dine produced, supersymmetric DM.  Specifically we take the DM candidate to be the lightest mass eigenstate of some linear combination of active sneutrinos, $\tilde{\nu}$ and an additional sterile sneutrino, $\tilde{N}$.  Such mixed sneutrino DM is discussed, for example, in \cite{sneutrino0,sneutrino1,sneutrino2,sneutrino3}.  This DM particle carries lepton number of +1 and based on the initial vev for the DM field, there can be an initial asymmetry between the number of DM particles and antiparticles, making DM annihilation negligible today.  

Since the DM field, $\chi$, is a superposition of $\tilde{\nu}$ and $\tilde{N}$ we can write
\begin{align}
\tilde{\nu} = \psi\cos\theta + \chi\sin\theta ~\text{  and} \\
\tilde{N} = -\psi\sin\theta + \chi\cos\theta~,
\end{align}
where $\psi$ is the heavier mass eigenstate.  Note that for $\sin\theta \lesssim 0.27$, $m_{\chi}$ is unconstrained by the invisible Z width \cite{Zwidth}.

$\tilde{N}$ is a weak isosinglet with only gravitational interactions and $\tilde{\nu}$ is in a weak doublet.  We take the dominant interactions of $\chi$ to be with the weak gauge bosons (i.e. its couplings via the superpotential are negligible and its coupling to $\psi$ is kinematically suppressed).  We can find $\chi$'s weak interactions with gauge bosons by using the covariant derivative from the $SU(2)_L~\times~U(1)_Y$ symmetry of the standard model, so that kinetic terms for $\tilde{\nu}$ and $\tilde{N}$ in the Lagrangian are given by
\begin{equation}
\label{lagrangian}
\mathcal L_{kin} = -D^{\mu}\tilde{\nu}^{\dagger}D_{\mu}\tilde{\nu}-D^{\mu}\tilde{N}^{\dagger}D_{\mu}\tilde{N}~,
\end{equation}
where  $ D_{\mu}\tilde{\nu} = \partial_{\mu}\tilde{\nu} - i(g_1B_{\mu}Y+g_2A^a_{\mu}T^a)\tilde{\nu}$ and $D_{\mu}\tilde{N} = \partial_{\mu}\tilde{N}$.  The relation of the $A_{\mu}^a$ and $B_{\mu}$ gauge bosons to the standard model photon $A_{\mu}$, the $Z^0$, and the $W^{\pm}$ bosons is
\begin{equation}
\begin{tabular}{c}
$A_{\mu}^1 = \frac{1}{\sqrt{2}}(W_{\mu}^-+W_{\mu}^+)$ \\
$A_{\mu}^2 = \frac{1}{\sqrt{2}i}(W_{\mu}^--W_{\mu}^+)$ \\
$A_{\mu}^3 = Z_{\mu}^0\cos\theta_w+A_{\mu}\sin\theta_w$ \\
$B_{\mu} = A_{\mu}\cos\theta_w-Z_{\mu}^0\sin\theta_w~,$ \\
\end{tabular}
\end{equation}
where $\tan\theta_w=\frac{g_1}{g_2}$ and $e = g_2\sin\theta_w$, where $e$ is the magnitude of the electron charge and $\theta_w$ is the weak mixing angle. Replacing $\tilde{N}$ and $\tilde{\nu}$ with $\chi$ and $\psi$, we find that the interaction Lagrangian for $\chi$ is given by
\begin{align}
\mathcal L_{int}= &\frac{i \sin^2\theta}{2}\sqrt{g_1^2+g_2^2}Z_{\mu}^0\left[\partial^{\mu}\chi^{\dagger}\chi - \chi^{\dagger}\partial^{\mu}\chi\right] \\
& - \frac{\sin^2\theta}{2}g_2^2\chi^{\dagger}W_{\mu}^+W^{-\mu}\chi - \frac{\sin^2\theta}{4}(g_1^2+g_2^2)\chi^{\dagger}Z_{\mu}^0Z^{0\mu}\chi~. \nonumber
\end{align}

If the the four-momentum transfer squared in a DM-$Z/W^{\pm}$ interaction is less than a few GeV, then the first term in the above interaction Lagrangian is dominant.  We may also integrate out the $Z$ in the remaining term, giving us the generic effective Lagrangian form that we had in (\ref{Lagrangian}).  In this case, the effective coupling constant is given by
\begin{equation}
\tilde{G} = \frac{e^2\sin^2\theta c_V}{2M^2_Z\sin^2\theta_w\cos^2\theta_w}~,
\end{equation}
where $c_V = \frac{1}{4}$ for a neutron and $c_V = -\frac{1}{4} + \sin^2\theta_w$ for an electron and all the results of the previous sections can be applied.  



\bibliographystyle{apsrev}
\bibliography{dark_matter_references}

\begin{thebibliography}{67}
\bibitem[{DMd()}]{DMdensity}
\bibinfo{note}{Many pulsars have been observed in globular clusters
  \cite{pulsarsgc} which can have DM densities as large as $\rho_{\chi}\sim
  800$ GeV/cm${}^3$ (or perhaps larger) as predicted for the globular cluster
  M4 \cite{M4} however the presence of such large DM densities in globular
  clusters is uncertain--see also \cite{lessDM}}.
\end{thebibliography}

\end{document}